\input epsf
\documentstyle[11pt,paspconf]{article}
\begin{document}

\title{Cosmic Glows}
\author{Douglas Scott}
\affil{Department of Physics and Astronomy, University of British Columbia,
          Vancouver, B.C. V6T 1Z1\ \  Canada}

\begin{abstract}
This is the obligatory Cosmic Microwave Background review.  I discuss
the current status of CMB anisotropies, together with some
points on the related topic of the Far-Infrared Background.
We have already learned a number of important things from CMB
anisotropies.  Models which are in good shape have: approximately flat
geometry; cold dark-matter, plus something like a cosmological constant;
roughly scale invariant adiabatic fluctuations; and close to Gaussian
statistics.
The constraints from the CMB are beginning to
be comparable to those from other cosmological measurements.
With a wealth of new data coming in, it is expected that CMB anisotropies
will soon provide the most stringent limits on a combination of
fundamental cosmological parameters.

\noindent
{\it Keywords:} CMB -- FIB -- Monty Python
\keywords{Cosmic Microwave Background, Far-Infrared Background, Monty Python}
\end{abstract}

\section{`Always Look on the Bright Side of Life'}

Much of the matter in the Universe is in a form which is not very luminous.
There is additional evidence that majority of the cosmological density is
contained in some even more mysterious Dark Energy.  However, we learn
most about the Universe by studying the properties of its radiation, which
is, of course, much easier to see than the dark stuff.

Fig.~1 gives an overview of information on extragalactic background
radiation in the Universe.  $\nu I_\nu$ is plotted so that it is possible
to read off the relative contributions to total energy density --
the Cosmic Microwave Background (CMB) is by far
the dominant background.
On the figure, the next biggest background
-- almost two orders of magnitude
down in energy  contribution -- is the Far-Infrared Background (FIB),
which is believed to be produced by distant, dusty, star-forming galaxies.
A little below that is the near-IR/optical background, coming from the sum
of the emission of all the stars in all the galaxies we can observe.  Then
much lower are the X-ray and $\gamma$-ray backgrounds,
which come predominantly
from active galactic nuclei.

\begin{figure}
\begin{center}
\epsfxsize=13cm \epsfbox{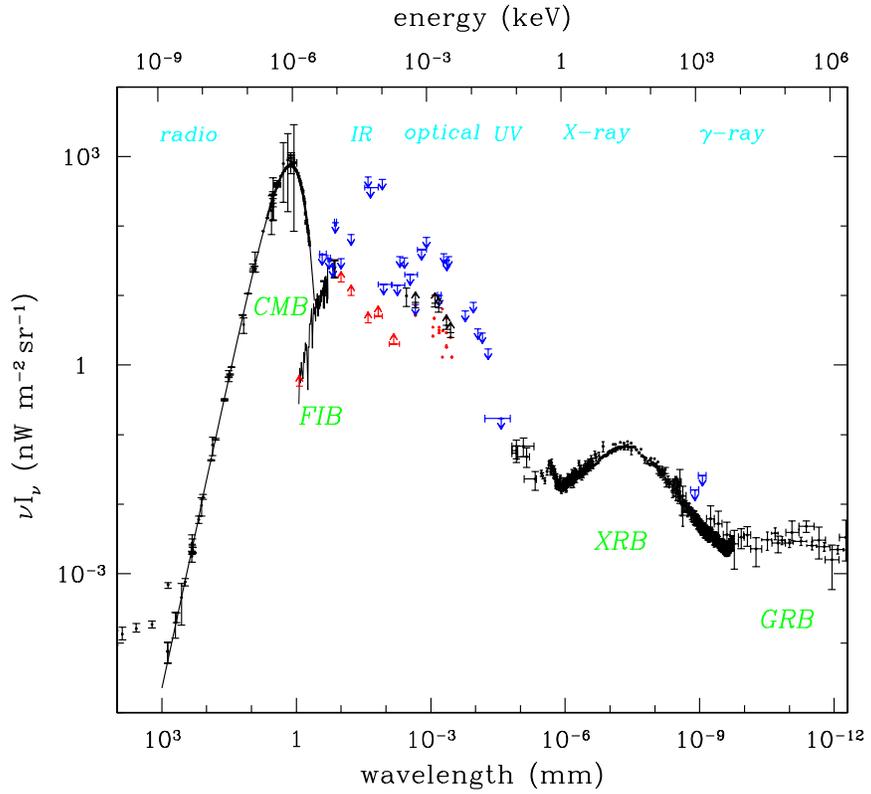}
\end{center}
\caption{A compilation of recent constraints on extragalactic diffuse
background radiation.  In terms of total energy the CMB dominates, with the
Far-Infrared and Optical Backgrounds about a factor of 100 lower.
These data are based upon the older compilation of Ressell \&
Turner~(1990), supplemented with more recent data from:
Smoot~(1997) for the CMB; Lagache et al.~(1999) and
Hauser et al.~(1998) for the FIB;
Leinert et al.~(1998) for a near-IR to near-UV compilation;
Dwek \& Arendt~(1998) for the near-IR;
Pozzetti et al.~(1998) and Madau \& Pozzetti (1999) for the optical;
Miyaji et al.~(1998) and Gendreau et al.~(1995) for the X-ray;
and Sreekumar et al.~(1998),
Kappadath et al.~(1999), Weidenspointner et al.~(1999) and Watanabe
et al.~(1997) for the $\gamma$-ray.}
\end{figure}

We can learn a great deal by studying the photons contained in these
background radiations.  The CMB spectral shape is spectacularly well-fit
by a blackbody (Fixsen et al.~1996, Smoot~1997),
over more than 4 decades in frequency.
The current best estimate of the CMB temperature is
\begin{equation}
T_0 = 2.725 \pm 0.001\ {\rm Kelvin}
\end{equation}
(Mather et al.~1999).
For a recent discussion of what can be learned from the CMB spectral
distortions, and related effects, see Halpern \& Scott (1999).
Since the Universe recombined at $z\,{\simeq}\,1000$
(see Seager, Sasselov \& Scott~1999 for an update)
then that is when the CMB photons last interacted with matter.  Hence
a study of the anisotropies in the CMB tells us about inhomogeneities 
at $z\,{\sim}\,1000$.  I will focus on this topic for most of the rest of
this article (see White, Scott \& Silk~1994, Scott, Silk \&
White~1995 and Smoot \& Scott~1998 for older but more comprehensive reviews).

\section{`And Now for Something Completely Different'}

Counts of galaxies, from the SCUBA instrument on the James Clerk Maxwell
Telescope, amount to almost $10^4\,{\rm deg}^{-2}$ down to $1\,$mJy
(e.g.~Hughes et al.~1998, Barger et al.~1999, Blain et al.~1999,
Chapman et al.~1999, Lilly et al.~1999).
This accounts for at least half of the background at these wavelengths,
and {\it all\/} of the
background can be explained by extrapolating the counts down to, say,
$0.5\,$mJy.  Summing up the contributions
of these SCUBA sources provides a direct measurement of the FIB at
$850\,\mu$m.  However, this is a wavelength where in fact the diffuse
measurement, obtained from FIRAS
data (Fixsen et al.~1998), is not so well constrained.

Fig.~2 shows a compilation of counts at a range of infra-red wavelengths.
The solid line is a prediction of a model of galaxy formation and evolution
by Guiderdoni et al.~(1998).  Galaxies of different type, luminosity and
redshift will contribute variable amounts to the background at each
wavelength.
The FIB appears to peak at about $200\,\mu$m (Hauser et al.~1999) -- since
this is a wavelength range which is essentially impossible from the
ground, there is little direct galaxy data.  Reasonable extrapolations of
the spectra of SCUBA galaxies give some constraints at ${\sim}\,200\,\mu$m,
but SCUBA-bright galaxies probably have a bias towards being at higher
redshifts than those contributing to the bulk of the FIB at its peak.
The best {\it direct\/} means currently for investigating the galaxies which
comprise the FIB is through follow-up of sources from the 
Far InfraRed BACKground survey (see e.g.~Dole 1999).
The FIRBACK survey
represents the deepest extensive
$170\,\mu$m images obtained by the ISO satellite, and detected the brightest
galaxies which comprise ${\sim}\,10$\% of the background.  The
combination of $170\,\mu$m from ISO and $850\,\mu$m with SCUBA is particularly
interesting as a redshift discriminator (Scott et al.~1999).  Follow-up
at a wide range of wavelengths should soon give us a fairly complete
picture of at least these brightest contributors to the FIB.

\begin{figure}
\epsfxsize=12cm \epsfbox{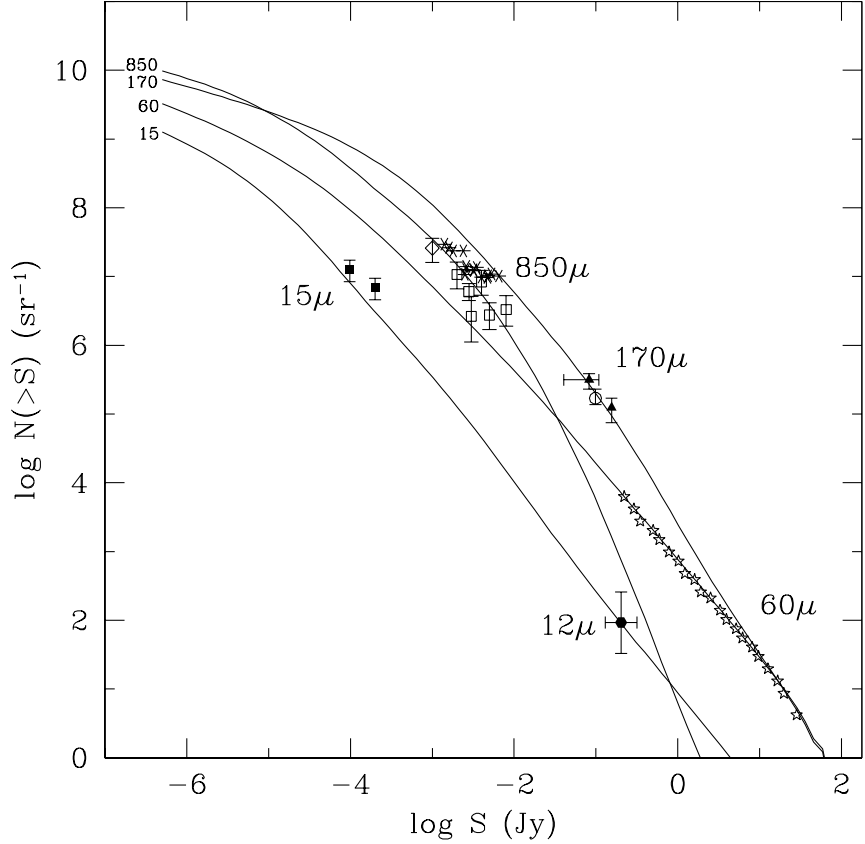}
\caption{Source counts at $15\,\mu$m, $60\,\mu$m,
$170\,\mu$m and $850\,\mu$m (based on figure~17 of Guiderdoni et al.~1998).
The data are from: IRAS -- $12\,\mu$m filled
(hexagon, Rush et al.~1993)
and $60\,\mu$m (stars, Lonsdale et al.~1990); ISO -- $15\,\mu$m (filled
isquares , Oliver et al.~1997) and
$170\,\mu$m open hexagon (Kawara et al.~1998)  and solid
triangles (FIRBACK project, Puget et al.~1998); and SCUBA
-- (open squares, compilation from Scott \& White 1998; asterisks,
Chapman et al.~1999; diamond, Blain et al.~1999).  The solid lines are
predictions from a particular semi-analytic model for
of galaxy formation and evolution (`model E' of
Guiderdoni et al.~1998) -- more detailed information about galaxy types,
redshifts and luminosities should discriminate between specific models.}
\end{figure}

Other experiments are also focussed on this waveband.  Although it is
difficult to reach wavelengths much short than $850\,\mu$m from
even the highest astronomical observatories, it is relatively
straightforward from balloons.  A new sub-millimetre
long-duration balloon experiment, the Balloon-borne Large-aperture
Sub-millimeter Telescope (BLAST), will
operate down to $350\,\mu$m, with the goal of surveying
distant star-forming galaxies (as well as making high-resolution maps
of galactic emission, and searching for Sunyaev-Zel'dovich clusters).
BLAST will use a large array of bolometric
detectors, similar to those being developed for the FIRST satellite.
Operating at shorter wavelengths, it will nicely complement BOLOCAM
(soon to be installed at the CSO) and the upgraded version of SCUBA.
With all of these new instruments coming on-line,
we will soon have a much more complete understanding of the galaxies
which make up this second-highest cosmic background.

\section{`And the Number of the Counting Shall be Three'}

As well as investigating the diffuse backgrounds (the DC or monopole terms,
$\ell\,{=}\,0$)
we can also look at variations over the sky.
The next highest order moment is the dipole, $\ell\,{=}\,1$,
which is described by $(2\ell+1)\,{=}\,3$ independent numbers.  The CMB dipole
is firmly believed to be `extrinsic', i.e.~caused by our motion
through the sea of CMB photons, rather than being an intrinsic anisotropy
on the sky (which would be expected to be of roughly the same amplitude
as the quadrupole, which is about 100 times smaller).
In principle the dipole should exist in {\it all\/}
of the backgrounds (although local galaxies will complicate matters at
many wavelengths) -- indeed there
is some evidence in the XRB (Scharfe et al.~1999).
But only the CMB measurements (Fixsen et al.~1996, Lineweaver et al.~1996)
are precise enough to accurately determine our local motion:
\begin{eqnarray}
v_\odot&=&371\pm0.5\,{\rm km}\,{\rm s}^{-1}\\
{\rm towards}\quad
(\alpha,\delta) &=& (11\fh20 \pm 0\fh01, -7.22^{\circ}
\pm 0\fdg08)\nonumber\\
{\rm or}\quad
(\ell,b) &=& (264\fdg31\pm0\fdg17, 48\fdg05\pm0\fdg10).
\nonumber
\end{eqnarray}
This implies a velocity for the Local Group relative to the CMB of
$v_{\rm LG} = 627 \pm 22\,{\rm km}\,{\rm s}^{-1}$ toward
$(\ell,b) 
= (276^{\circ} \pm  3^{\circ}, 30^{\circ} \pm 3^{\circ})$,
where most of the error 
comes from uncertainty in the velocity of the solar
system relative to the Local Group (see Courteau \& van den Bergh~1999 for
an update).  Our motion is just one realization of
a statistical velocity field -- other galaxies in the Universe will
see different dipoles, and the frame with {\it no\/}
dipole defines, in a sense, the absolute rest frame of the Universe.
The relationship between 
the local mass field and these large scale velocity fields is the subject
of most of the rest of this volume.

\section{`Here Comes Another One'}

Although the dipole may be the most directly relevant multipole for Cosmic
Flows, there are of course many more multipoles: the quadrupole, the
octupole, not to mention the hexadecapole, and others besides which I don't
know the names of.  The final maps produced by the {\sl Planck} satellite,
for example,
are expected to contain perhaps 3{,}000{,}000 modes out to the multipoles
where the noise starts to dominate the signal.

These higher order moments of the background anisotropies tell us about
correlations on the sky.  For backgrounds which are produced by point
sources (likely to be true for everything except the CMB), these
anisotropies come from a combination of Poisson fluctuations and the
intrinsic clustering of the sources (see for example Scott \& White~1999
and Haiman \& Knox~1999).  Such fluctuations have already been seen in
both the FIB (Lagache \& Puget~1999) and the XRB (Treyer et al.~1998).
And of course clustering of the objects which make up the optical
background has been well-known for decades.
It seems likely that studies of clustering in the FIB will flourish as
a tool to learn more about the properties of the populations comprising it.

But meanwhile the anisotropies in the CMB are already providing a wealth
of cosmological information.
Soon after the detection of CMB anisotropy by the {\sl COBE\/} satellite
(Smoot et al.~1992) it became clear that anisotropy measurement
was easily within reach of state-of-the-art detectors.  In addition
careful theoretical calculation showed that precise measurements of the
anisotropy power spectrum would provide detailed information about fundamental
cosmological parameters (e.g.~Bond et al.~1994, Jungman et al.~1995,
Hu et al.~1995, Seljak \& Zaldarriaga~1996).

\begin{figure}
\begin{center}
\epsfxsize=12.5cm \epsfbox{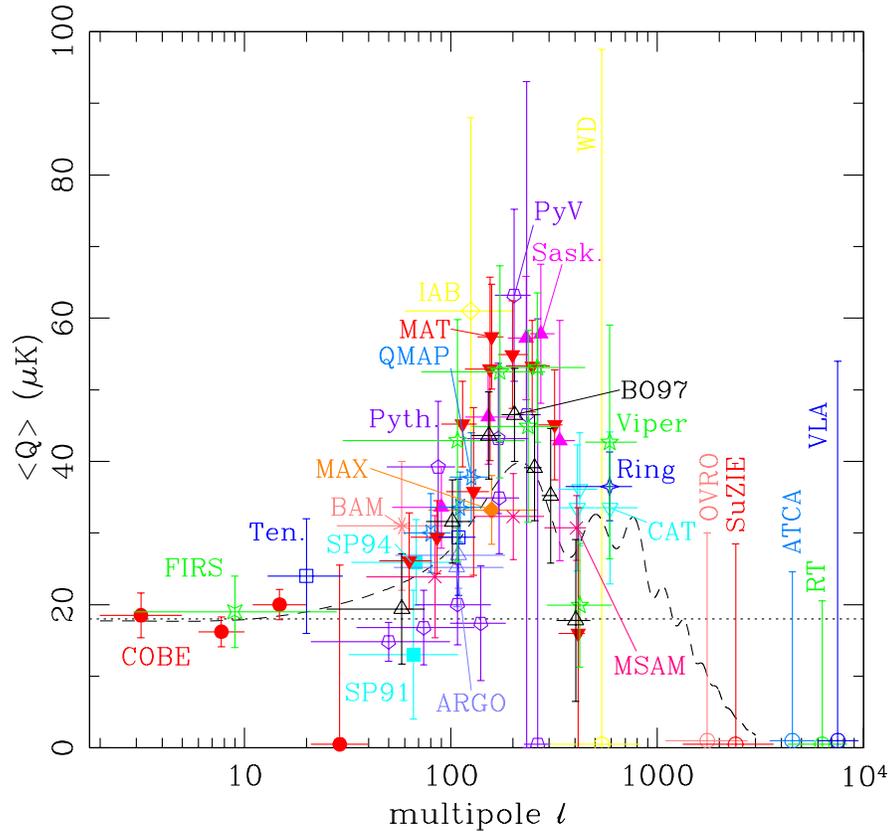}
\end{center}
\caption{Most of the CMB anisotropy
experiments published to date.  See Smoot \& Scott
(1997)  and Halpern \& Scott (1999) for full references,
supplemented with more recent
results from: OVRO Ring (Leitch et al.~1999); QMAP (de Oliveira-Costa
et al.~1998); MAT (a.k.a.~TOCO, Torbet et al.~1999 and Miller et al.~1999);
CAT (Baker et al.~1999); Python V (Coble et al.~1999);
Viper (Peterson et al.~1999); and BOOMERanG NA (a.k.a.~BOOM97, Mauskopf
et al.~1999).  The error
bars (these are $1\sigma$ except for the upper limits which are 95\%)
have generally been symmetrised for clarity, and calibration
uncertainties are included in most cases. The horizontal bars represent
the widths of the experimental window functions.  The dotted line is the
flat power spectrum which best fits the {\sl COBE\/} data alone.  The dashed
curve is the prediction from the vanilla-flavoured
standard Cold Dark Matter model.}
\end{figure}

There have been around 20 separate experiments which have detected
temperature fluctuations which are most likely to be primordial.  These
are summarized in Fig.~3.
Here the $x$-axis is the spherical harmonic multipole, $\ell$.  The temperature
fluctuation field on the sky, excluding the monopole and dipole,
can be decomposed into an orthogonal set of
modes:
\begin{equation}
{\Delta T\over T}\left(\theta,\phi\right)
 = \sum_{\ell=2}^\infty \sum_{m=-\ell}^{+\ell}
 a_{\ell m} Y_{\ell m}(\theta,\phi).
\end{equation}
Since there is no preferred direction on the sky
(e.g.~Bunn \& Scott~1999) the individual $m$s are irrelevant, and so the
important information is contained in the power spectrum
\begin{equation}
C_\ell \equiv \left\langle \left| a_{\ell m} \right|^2 \right\rangle.
\end{equation}
Indeed if the perturbations are Gaussian, then this contains {\it all\/} the
information.
The conventional amplitude of the quadrupole is given as
\begin{equation}
{Q^2\over T_0^2} \equiv {\sum_m \left| a_{\ell m} \right|^2\over4\pi}
= {5C_2\over 4\pi}.
\end{equation}
A `flat' spectrum means one in which
$\ell(\ell+1)C_\ell={\rm constant}$, and we can therefore define that
constant in terms of the expectation value for the
equivalent quadrupole $\left\langle Q_{\rm flat}\right\rangle$ -- which
is what is plotted as the $y$-axis in Fig.~3.

Each experiment quotes one
(or in the best cases several) measures of power over a range of multipoles,
and these can be quoted as `band powers' or equivalent amplitudes of a flat
power spectrum through some `window function'.  The horizontal bars on the
points are an indication of the widths of these window functions (see
White \& Srednicki~1996 and Knox~1999 for more details).

More complete references for the experiments can be found in,
e.g.~Smoot \& Scott~(1997), Halpern \& Scott~(1999), Scott~(1999).
And for more details on each experiment start with
{\tt http://www.astro.ubc.ca/people/scott/cmb.html},
or other similar web-pages.

\section{`Amongst our Weaponry are such Elements as \dots'}

The list of the things which we have already learned
from the CMB is probably larger than expected.
Several facts can immediately be gleaned from Fig.~3:
\begin{itemize}
\item The plot has become very crowded!
\item The overall detection of
anisotropy is at the ${\simeq}\,40\sigma$ level
\item
A flat power spectrum (horizontal dotted line) is a bad fit, at
about the $20\sigma$ level
\item
There is clear evidence for a peak at $\ell\sim200$
\end{itemize}
One thing to add is that despite the appearance of scatter in the plot,
smooth curves exist which provide a perfectly good fit to all the
results (with only one or two exceptions).
In order to see the wood for the trees, I have provided a binned version
of the data in Fig.~4.  

We can learn about the large-scale properties of the Universe from a variety
of different techniques: cosmic flows; distant supernovae;
galaxy clustering; ages of globular
clusters; measurements of $H_0$; direct estimates of $\Omega_0$; cluster
abundance; cluster baryon fraction; Big Bang nucleosynthesis; Lyman
forest fluctuations; and other things I've forgotten.
We are now at the point where the CMB anisotropies are providing constraints
at least as good as from these other approaches.  CMB anisotropies are an
ideal tool for attacking
cosmological parameters, since they probe linear theory
on large scales.  There is a great deal more information to obtain before
the CMB bottoms-out (due to the cosmic variance limit), and no reason to
be particularly concerned that any unavoidable systematic effect will
be a show-stopper.  Certainly there is a great deal to learn from other
approaches (e.g.~Cosmic Flows)
which provide complementary information, but it seems clear that
soon there will be extraordinarily large amounts of cosmological data
collected from the CMB sky.

\begin{figure}
\epsfxsize=12.5cm \epsfbox{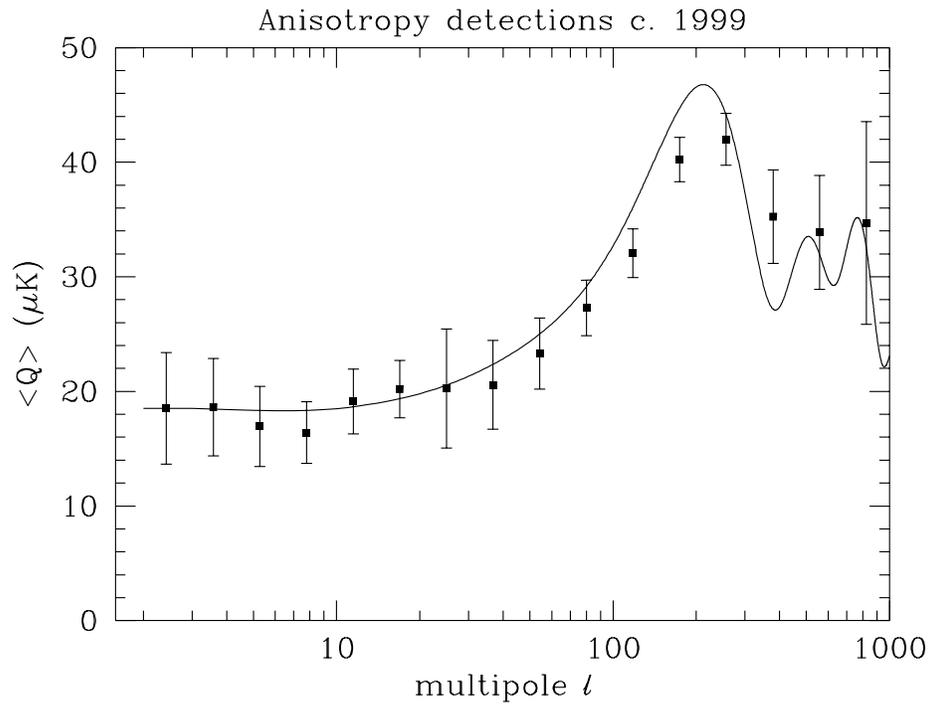}
\caption{The result of binning the available data (from Fig.~3 for example).
More precisely, what was done was to split the multipoles into
16 bins between $\ell\,{=}\,2$ and $\ell\,{=}\,1000$,
and to weight each experiment
by the fraction of the window in each bin.  The precise height of the
peak depends to some extent on the choice of bins, on details of the
window functions used, and on the weights given to individual experiments.
The points here are {\it not\/} un-correlated, but provide a reasonable visual
impression of the current data -- more sophisticated treatments
(e.g.~Bond, Jaffe \& Knox~1998) give similar results.
The solid line is a $\Lambda$-dominated CDM model, with parameters which
are consistent with most current cosmological constraints.}
\end{figure}

The first thing we learn from Figs.~3 and 4 is that
our basic paradigm -- to describe the large scale properties of the Universe,
and the formation of structure within it -- are in good shape.  The
prediction from the `straw man' model, standard Cold
Dark Matter (sCDM) is shown by the dashed line in Fig.~3.
This model (which contains parameters which
are all fixed at very round numbers) is hardly the best fit, but it
does have the right general character.  And
it is easy to find models which fit the data much better, by tuning some
of those parameters.

More detailed inferences from Fig.~4 can be listed as follows
(for more details refer to the review article by Lawrence et al.~1999):
\begin{itemize}
\item Gravitational instability in a dark matter dominated universe
grew today's structure
\item The Universe remained neutral until $z\,{\la}\,50$
\item The CMB power spectrum peaks at
$150\,{\la}\,\ell\,{\la}\,350$
\item There are some (weak) constraints on particle physics at
$z\,{\sim}\,1000$
\item The large-scale structure of spacetime appears to be simple
\end{itemize}

I will not discuss most of these in detail, but
let me return to the issue of the main CMB acoustic peak.
Since the standard
CDM model has its main peak at $\ell\,{\simeq}\,220$, and it is pushed to
smaller scales in open models,
then the data prefer $\Omega_{\rm tot}$ close to unity, and
certainly bigger than the ${\sim}\,0.3$ implied by various dynamical studies.
Rigorous analyses of the anisotropy data (e.g.~Dodelson \& Knox~1999,
Melchiorri et al.~1999) arrive at similar conclusions.
The CMB thus provides the strongest evidence that the Universe is flat
(or at least, nowhere near as open as the amount of dynamically detected
dark matter would suggest).

The height of the peak is somewhat greater than predicted for sCDM, but
entirely consistent with several variants.
Currently popular models with a cosmological
constant tend to provide perfectly good fits to the CMB.
The curve
plotted in Fig.~4, shows one such flat model with $\Omega_\Lambda=0.6$
and a Hubble constant of $70\,{\rm km}\,{\rm s}^{-1}{\rm Mpc}^{-1}$.

Since the height of the first peak depends on a combination of
parameters, then exactly what quantities are constrained
depends on the parameter space being searched, as well as on the choice
of additional constraints.  Currently it is possible to constrain
the matter density $\Omega_{\rm M}$ to ${\sim}\,\pm0.1$ from the peak height,
but that depends sensitively on the assumptions used.  All this is
expected to change as better data come in.

All of this parameter estimation depends on having the correct family
of models to test.  In fact it is clear that
that models with adiabatic-type perturbations
(i.e.~where you perturb the matter and radiation at the same time in order
to keep the entropy fixed) have the right kind of character to fit
the data.
On the other hand isocurvature-type models (where the matter and radiation
get equal and opposite perturbations, so that the local curvature is
unperturbed) tend to look poor -- generically they have a `shoulder'
rather than a first peak, and then the highest peak is at much smaller
scale (see e.g.~Hu, Spergel \& White~1997).
While there are some loop-holes, it seems difficult to get
isocurvature models to fit the current data.  The basic thing to take away
from Fig.~4 then is that adiabatic models (with roughly scale-invariant
initial conditions) are in good shape, and that within this class of 
models the CMB data are beginning to put constraints on parameters.

\section{`Is this the Right Room for an Argument?'}

The most contentious conclusion from the current CMB data concerns the
implications from the mechanism that generated the perturbations.
Most people working in the field which is sometimes referred to as
`CMB phenomenology' are currently struggling with the same question, in one
form or another: does the CMB data prove inflation?
It seems clear that we are now in a position to say {\it something\/}
beyond the Big Bang paradigm.  The CMB led us to accept that the Universe
used to be hotter and denser, and more recently to the conclusion that
structure built
up through gravitational instability.  Now it appears that we are learning
something further, something about the origin of the perturbations themselves.
But just exactly {\it what\/} that next step is, and how to phrase it, is
altogether less clear.  For lack of anything better, let me present my own
current belief, which I challenge anyone to disagree with:
\begin{itemize}
\item Something like Inflation is something like proven
\end{itemize}

The only causal
way we know of to have adiabatic fluctuations on apparently acausal scales
is to have the scale factor accelerate
at some time in the early history of the Universe
(e.g.~Liddle~1995).  And we
can argue about whether something that achieves the same end result is just
isomorphic to inflation, even if interpretted differently.  Here `inflation'
does not necessarily carry with it the extra baggage of an inflaton
potential etc. -- although hopefully the connection with particle physics
would follow later.

It used to be that discussions of inflation focussed on the number of
e-foldings required to solve horizon, flatness, entropy and monopole
problems.  However, at the present time the paramount concern is making the
density perturbations -- and inflation gives you a mechanism to do
that, for free.  It appears that we are learning that the Universe has
inflation-like `initial conditions'.  Time will tell whether that means
that the Universe was once dominated by some vacuum energy density, and
whether we can learn details about particle physics at ultra-high energies.

\section{`This is an Ex-Parrot'}

Let me continue the contentious theme by saying a little something about the
main competition for inflation --
any one of various
field ordering mechanisms or topological defect models.  Generically these
give larger CMB anisotropies, from the so-called integrated Sachs-Wolfe
effect, for the same density perturbation amplitudes.  This
is basically because of their similarity to isocurvature models; adiabatic
(i.e.~inflationary) models, on the other hand, give the correct value to a
factor of 2, without really trying.  The power spectra of
galaxy perturbations, or even the underlying dark matter fluctuations, are
notoriously complicated to calculate in defect models
-- nevertheless there seems little hope
that the observed power spectrum can be easily reproduced in these
sorts of models.  Moreover, there now seems to be some consensus in the view
that generic defect models produce either one (broad) peak
in a place which gives a poor fit to current data,
or perhaps even no peak at all (Pen, Seljak
\& Turok~1997, Albrecht et al.~1997, Allen et al.~1997).

The status of defects vs. the Universe can be summarized in the following
three points.
Defect models tend to give:
the wrong matter power spectrum;
the wrong CMB power spectrum;
and the wrong normalization of matter relative to CMB.
But apart from that, these models seem to work fine!

Of course there is some motivation for working on such models simply
because they {\it cool\/} -- and indeed they may yet be
important in other purposes in the early Universe --
but at this point they seem to hold little promise as a method of
forming structure.

\section{`This Theory which Belongs to me is as Follows'}

If we are close to proving something like inflation, then that may only be
because the only serious competitor, defects, seems in such bad shape.
It may be that we are just lacking in imaginative enough ideas, just waiting
for a better theory to come along.
So certainly it is worth
investigating other possibilities,
at least until such time as inflation has been more directly tested.
While the current suite of defect models do not look very promising, there
is always the possibility of a more attractive, related model lying around
corner.

New ideas from particle physics also have the potential for providing
different mechanisms for structure formation.  Exactly what will come out
of string theory, large extra dimensions and broken Lorentz invariance
remains to be seen.  It will be interesting to see how generic the basic
inflationary predictions are, and whether new twists on high energy
physics carry with them new testable predictions.

\section{`That's the Machine that Goes ``Ping'''}

That the field of CMB anisotropies is progressing rapidly is largely due
to the experimental efforts.  These are extremely difficult measurements
to make, and to do so has required impressive developments in detector
technology, as well as experimental strategies and data analysis methods.
More information about current and future CMB experiments can be found in
recent reviews (e.g.~Lawrence~1998, Halpern \& Scott~1999);
here I give only a very brief summary.

The next generation of CMB balloon experiments are expected to return data
of much higher quality (and quantity).  The newest results from the North
American test flight of BOOMERanG (`BOOM97', Mauskopf et al.~1999) are
certainly impressive, and others (such
as MSAM and MAXIMA) are eagerly awaited.  BOOMERanG `98 was the
first long-duration balloon flight, and by all accounts was staggeringly
successful -- BOOM98 seems likely to provide an enormous leap forward
in anisotropy measurements.  Three immediate questions are expected to be
addressed by this
new data-set: do the currently favoured $\Lambda$-dominated cosmologies
continue to be a good fit; what is the precise location of the first peak;
and is there any evidence for other peaks.  This last point is perhaps
the most important.  Detection of oscillations in the power spectrum, with
tight constraints on the peak spacings, will be a very firm test of the
inflationary paradigm (Hu \& White~1996).

The adiabatic, apparently acausal perturbations, generated during inflation,
give a series of peaks in the ratio $1\,{:}\,2\,{:}\,3\,{:}\,\cdots$
in $\ell$-space.  On the
other hand, causal, isocurvature perturbations naturally give rise to peaks
in the ratio $1\,{:}\,3\,{:}\,5\,{:}\,\cdots$.
So detection of a second peak at roughly half
the angular scale of the first, will be a very large step towards `proving
inflation'.  Failure to observe this will, of course, be even more exciting,
since it will demand an entirely new paradigm.

In the short term there are also at least three new interferometer
projects (e.g.~White et al.~1999): DASI at the South Pole; CBI in Chile;
and the
VSA in Tenerife.  All of these are nearing completion and should
produce data within the next year or two.

Another direction being pursued from the ground is CMB polarization --
see Staggs, Gundersen \& Church~1999 for experimental details and
Hu \& White~1997 for a
theory primer.  The CMB sky is naturally polarized
at the few percent level (the result of the quadrupole term in Compton
scattering together with a slightly anisotropic radiation field at
$z\,{\sim}\,1000$).  Measuring the ${\sim}\,\mu$K signals will be very
challenging, but can provide information beyond that contained in the
temperature anisotropies alone.  Since polarization is such a strong
prediction, it better {\it be\/}
there, otherwise our whole picture has to change!
Furthermore, we can more definitively separate any
gravity wave contribution in the CMB (if it is measurable,
Zibin, Scott \& White~1999),
thus limiting the energy scale of inflation.
Large-angle polarization can also constrain
the reionization epoch, and details of the polarization power spectrum
are a direct probe of physics around the time of last scattering.  This is
all in addition to the fact that polarization simply gives extra information
to better constrain parameters (and to break degeneracies between some
combinations of parameters).

Two satellite missions are currently planned to study the CMB from space,
where the whole sky can be imaged, far from the complicating effects of
the atmosphere.  The NASA Microwave Anisotropy Probe ({\sl MAP\/}
\footnote{{\tt http://map.gsfc.nasa.gov}})
is due for launch in November 2000.  It will travel to the Earth-Sun outer
Lagrange point, L2, where it will map the sky at 5 frequencies between
22 and $90\,$GHz, reaching to $\ell\,{\sim}\,800$ in the power spectrum.
The careful control of systematics possible with an
extended space mission means that {\sl MAP\/} should represent a very large
improvement over the data available from the Earth-based experiments.

\begin{figure}
\begin{center}
\epsfxsize=12.5cm \epsfbox{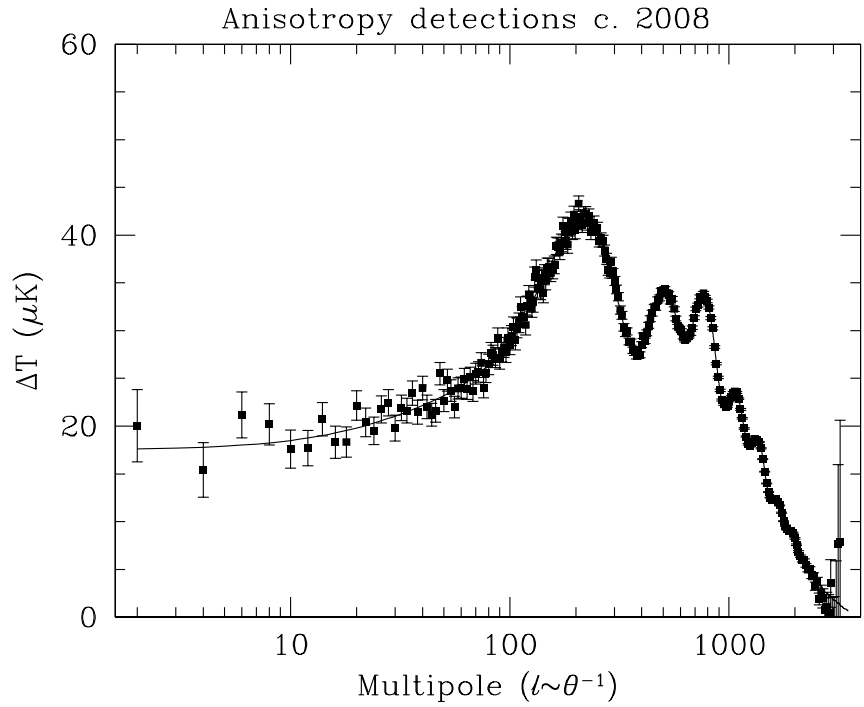}
\end{center}
\caption{
An estimate for how well the power spectrum might be measured by
{\sl Planck\/}.  This is a realization of a CDM power spectrum, assuming
the {\sl Planck} instrumental sensitivity over two thirds of the sky.
{\sl Planck\/} should supply us with essentially cosmic-variance limited
information on all the angular scales relevant to primary anisotropies, over
the full range of relevant frequencies.
}
\end{figure}

The ESA mission
{\sl Planck\/}\footnote{
{\tt http://astro.estec.esa.nl/SA-general/Projects/Planck/}}
can be thought of as the third generation CMB satellite,
mapping at 9 separate frequencies between 30 and $850\,$GHz, with both
radiometer and bolometer technologies, and measuring the $C_\ell$s
to beyond $\ell$ of 2000.  Thus {\sl Planck\/} is expected
to measure essentially {\it all\/} of the primordial CMB power
spectrum (see Fig.~5), and cover all the frequencies required to
measure and remove the foreground signals.  The {\sl Planck\/} data set
should enable cosmological parameters to be constrained with exquisite
precision -- or, to put it another way, the power spectrum should be
measured at a level of several million $\sigma$.
In addition {\sl Planck\/} will measure the polarization
(and cross-correlation with temperature) power spectra, providing even
more information.

Beyond {\sl MAP} and {\sl Planck} we might want to think about
polarization, about small angular scales and about a `CMB Deep Field'
(see Halpern \& Scott~1999 for discussion).  Diffuse `foreground' emission
(from the Galaxy as well as more distant sources) is also likely to be
studied more actively, as the ability to measure and identify these signals
develops.  Non-Gaussian
signals from higher-order effects at small-scales would be expected to show
up in fine-scale maps.

What will be the outcome of all this data?
Certainly cosmological parameters will be well constrained.
And definitely
some messy astrophysical details will be uncovered (in the foregrounds, as
well as through some weak processing effects occurring between
$z\,{=}\,0$ and 1000).  And {\it whatever\/} the basic paradigm, there will
surely be some clues to fundamental physics lurking in there, since the
CMB anisotropies provide the cleanest information about the initial
conditions and the largest scale properties of the Universe.

\section{`Exciting? No it's not. It's Dull, Dull, Dull'}

If one looks at what is currently known about cosmological parameters, then
depending on one's mood there are 2 possible
conclusions: {\it (1)\/} it is extremely
difficult to measure these things properly, and so the jury is still out;
or {\it (2)\/} we already have a fairly clear picture of the Universe.
I personally lean more towards the former (at least on Mondays, Wednesdays
and Fridays).  If we take the consensus, inflationary, $\Lambda$-dominated
models seriously, then we would conclude that all of the parameters
($\Omega_{\rm DM}$, $\Omega_{\rm B}$, $H_0$, $n$, $t_0$, negligible tensors,
reionization, non-adiabatic modes, no-Gaussianity, etc.) are known with
errors of perhaps 10--20\%.  It would be not only extremely surprising
if we were to find ourselves in that position, but also extremely
{\it boring\/}!

Let us fervently hope that the Universe is smarter than the smartest
theorist of the day, and is keeping something up its sleeve.  Otherwise
we will be reduced to measuring a bunch of model parameters to ever
greater precision.  Still, it is very hard to believe that `the end of
cosmology' is really in sight.  Rather than expecting {\sl MAP\/},
{\sl Planck\/} and the rest to be giving us 10 parameters to 1\%
accuracy, let us look forward to the surprises that are in store for us.

\section{`And Finally, Monsieur, a Wafer-Thin Mint'}

We are beginning to learn the answers to some fundamental questions,
using information contained in CMB
anisotropy data.  The CMB has already made it clear that gravitational
instability was sufficient to grow all the structure in the Universe.
Most recently it has been providing strong evidence that the Universe is
nearly flat.  Soon we can expect (through the existence or lack of
oscillations in the power spectrum) a direct test of the inflationary
scenario.  With future experiments, such as long-duration
balloons, interferometers, and the {\sl MAP\/} and
{\sl Planck\/} satellites, we should expect to learn vastly
more in the coming years about particle physics, cosmology and astrophysics.

\acknowledgments
I would like to thank my collaborators, in particular those involved
with {\sl Planck}, BLAST and various SCUBA-related projects.
Since this is a conference proceedings report, I
have shamelessly concentrated on my own work -- see the original papers
for more comprehensive references.  I am also grateful to the members of
CITA for their hospitality during the writing of this article.

\end{document}